# SlicerPET: A workflow based software module for PET/CT guided needle biopsy

*Dženan Zukić[1], Julien Finet[1], Emmanuel Wilson[2], Filip Banovac[3], Giuseppe Esposito[3], Kevin Cleary[2], and Andinet Enquobahrie[1]*

[1]Kitware Inc., Carrboro, NC, USA

[2]Sheikh Zayed Institute for Pediatric Surgical Innovation, Children's National Health System, Washington, DC, USA

[3]Department of Radiology, Georgetown University Medical Center, Washington, DC, USA

## Purpose

Biopsy is commonly used to confirm cancer diagnosis when radiologically indicated. Given the ability of PET to localize malignancies in heterogeneous tumors and tumors that do not have a CT correlate, PET/CT guided biopsy may improve the diagnostic yield of biopsies. To facilitate PET/CT guided needle biopsy, we developed a workflow that allows us to bring PET image guidance into the interventional CT suite. In this abstract, we present SlicerPET, a user-friendly workflow based module developed using open source software libraries to guide needle biopsy in the interventional suite.

## Methods

We have implemented a workflow-based module for PET/CT guided biopsy using the 3DSlicer framework [1]. 3DSlicer is a free and open-source extensible cross-platform toolkit for medical image segmentation, registration, visualization, and image-guided surgery. Slicer is in active use by the biomedical imaging research community as a vehicle to translate innovative algorithms into clinical research applications. The module is open source [2], corresponds with the clinical flow, and consists of the following steps:

1. **Data loading**: Allows a user to select the three volumes that are required for the guidance: respiratory-compensated PET, respiratory-compensated CT and interventional CT scans.
2. **Registration**: Register the respiratory-compensated CT with the interventional CT scan.

3. **Tracking:** Establish a connection between the tracking system and 3DSlicer using the Image Guided Surgery Toolkit (IGSTK).
4. **Tool calibration**: Determine the transformation between the tip and head of the biopsy needle if needed.
5. **Patient registration**: Register the interventional CT coordinate system to the patient coordinate system in the interventional suite using point-based registration.
6. **Needle path planning**: Identify entry and target point and generate plan for the needle path.
7. **Guidance**: Overlay a model of the needle and provide 3D visualization guidance for needle placement.

The workflow begins in the PET-CT suite where respiratory compensated PET-CT images are obtained. The patient will then move to the CT interventional suite for the biopsy. The motion-compensated CT from the PET-CT suite must then be registered with the CT from the interventional suite. The module allows for rigid body registration or deformable registration using a B-spline registration algorithm.

The biopsy needle has an electromagnetically tracked sensor (Aurora, Northern Digital, Waterloo, Canada) embedded in its tip to facilitate tracking. To connect this tracking device with our software, we use the open source library OpenIGTLink. To allow different tracking systems to be used, an optional tool calibration step is implemented. Because we are interested in tracking the tip of the needle, we need to calculate the linear transform from marker center to needle tip. To accomplish this reliably, we rely on PLUS library's PivotCalibration module [3]. The user places the needle's tip on a fixed point and pivots the handle until sufficiently many tracking points have been acquired for the algorithm to reliably estimate the transformation.

We also need to register the tracker coordinate system with the image coordinate system. Fiducials will be placed on the patient's body before the interventional CT acquisition, so we can do point to point registration. The needle tip is placed on a fiducial point, and then that point is selected in the interventional CT image using our software. Once at least four points have been registered this way, we can calculate the linear transform between tracker coordinates and image coordinates. We can also estimate reliability of the transform based on root mean squared error which is computed during transform matrix calculations.

Now the user can start planning the path for needle placement. The PET image layer is overlaid with the CT image and should assist the clinician in finding the area of increased metabolic activity. This area would then be selected on the user interface as the target point. The skin entry point is then selected and the proposed needle path is drawn between those two points over the CT image. This enables the clinicians to move the entry point until they are satisfied with a projected path which avoids blood vessels and other anatomy.

During the biopsy procedure, the needle will be tracked in real-time and updated on the virtual image. The clinicians will then use this display as guidance for the biopsy.

## Results

The entire system and workflow has been tested on a phantom. We use the computer-controlled, anthropomorphic respiring phantom shown on the left side of Fig. 1. This phantom has an embedded rib cage and an open thorax. Respiratory motion of the abdominal organs is induced by the motion of the diaphragm as air is pumped into and removed from the lung cavities. Respiration rate and volume are set using a computer controlled pump developed by the Georgetown group [4]. Within this thorax, we embed FDG-filled, sealable circular vials, which serve as targets and appear as areas of increased radiopharmaceutical activity in the PET images. The FDG is procured from the hospital supplier and comes as a single dose in a syringe. The dose is injected into a small, hollow, plastic sphere with a threaded seal. The sphere is sealed and placed inside a simulated liver made of foam, which has been placed in the anthropomorphic phantom.

We used a Siemens Somatom Definition AS_mCT 64 detector PET/CT machine. Single phase amplitude-gated acquisition protocol was used to minimize impact of respiratory motion. CT image size was 512x512x127 with 1.5x1.5x2.0 mm voxels. PET image size was 200x200x170 with 4.0x4.0x1.5 mm voxels. The interventional suite scanner is a Siemens Volume Zoom CT. The interventional image is 512x512x315 with 0.98x0.98x1.0 mm voxel size. All images were acquired at end expiration phase.

Image registration (CT to interventional CT) ended with a mutual information of 1.753 bits. Root mean square error of landmark registration in step 5 was 1.7189 mm.

The software is shown in Fig. 2.

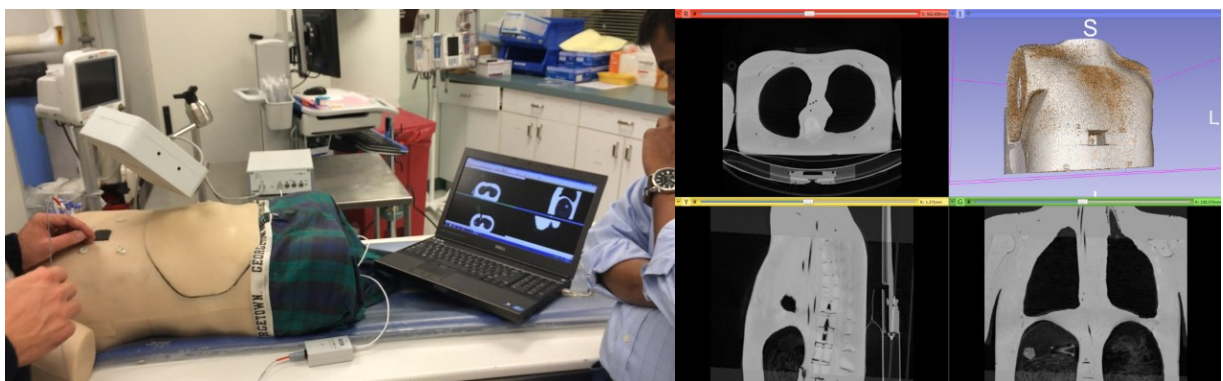

Fig. 1: Anthropomorphic phantom and CT-CT registration result

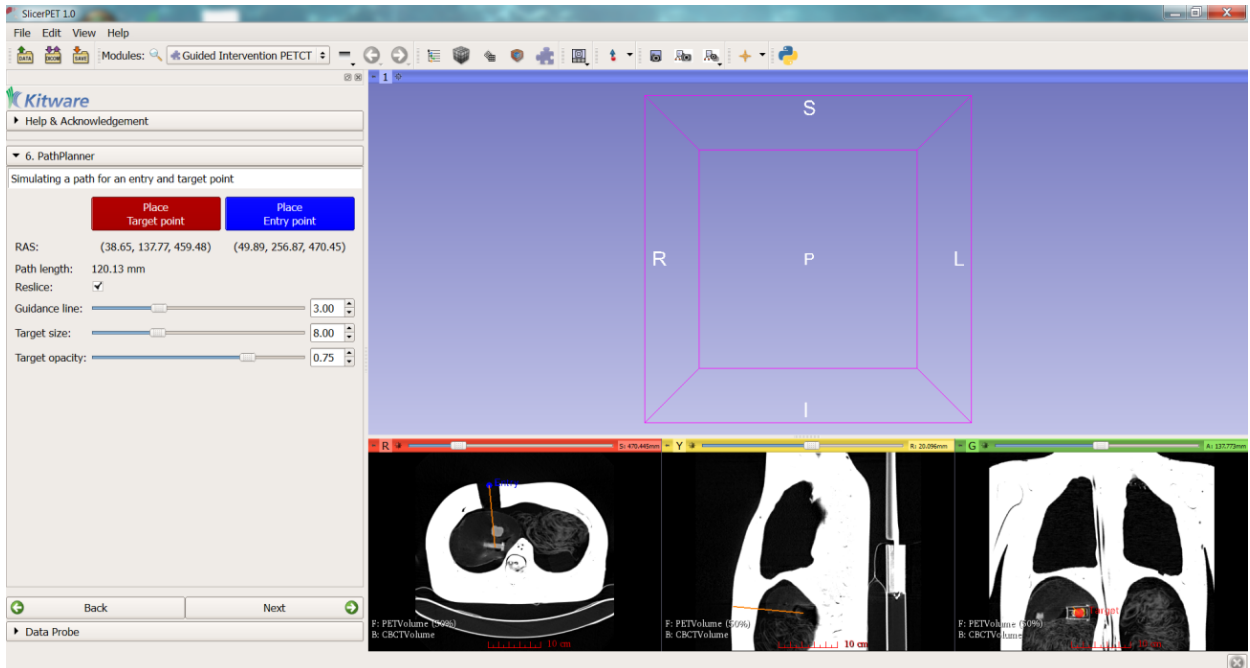

Fig. 2a: Path planning on fused PET-CT image

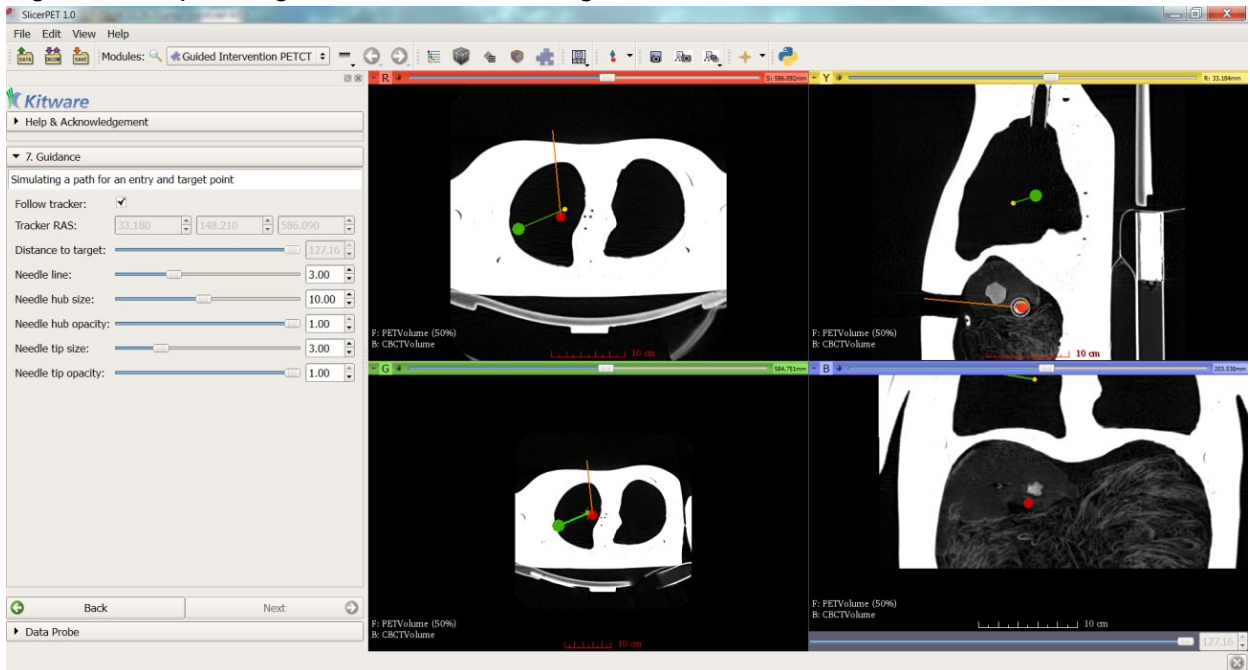

Fig. 2b: Needle guidance step

# Conclusion

We implemented an easy to use, step by step software module for PET-CT guided biopsy. The module was implemented in 3DSlicer and is available as open source software. Preliminary testing was done on an anthropomorphic phantom. The next step is to use the software in a clinical trial, for which we have already obtained IRB approval.